\documentclass[aps,pra,twocolumn,floatfix,superscriptaddress]{revtex4-1}

\usepackage{graphicx}        
\usepackage{amssymb}   
\usepackage{amsmath}
\usepackage{color,soul}
\usepackage{dcolumn}
\usepackage{bm}
\usepackage{mathtools}
\newcommand{\ket}[1]{{\vert}#1{\rangle}}
\newcommand{\bra}[1]{{\langle}#1{\vert}}
\begin{document}

\title{Remote polarization entanglement generation by local dephasing and frequency upconversion}

\author{S. Hamedani Raja}
\affiliation{Turku Centre for Quantum Physics, Department of Physics and Astronomy, University of Turku, FI-20014 Turun yliopisto, Finland}

\author{G. Karpat}
\affiliation{Faculdade de Ci\^encias, UNESP - Universidade Estadual Paulista, Bauru, SP, 17033-360, Brazil.}

\author{E.-M. Laine}
\affiliation{Turku Centre for Quantum Physics, Department of Physics and Astronomy, University of Turku, FI-20014 Turun yliopisto, Finland}

\author{S. Maniscalco}
\affiliation{Turku Centre for Quantum Physics, Department of Physics and Astronomy, University of Turku, FI-20014 Turun yliopisto, Finland}
\affiliation{Centre for Quantum Engineering, Department of Applied Physics, School of Science, Aalto University, P.O. Box 11000, FIN-00076 Aalto, Finland}

\author{J. Piilo}
\email{jyrki.piilo@utu.fi}
\affiliation{Turku Centre for Quantum Physics, Department of Physics and Astronomy, University of Turku, FI-20014 Turun yliopisto, Finland}

\author{C.-F. Li}
\email{cfli@ustc.edu.cn}
\affiliation{CAS Key Laboratory of Quantum Information, University of Science and Technology of China, Hefei, 230026, China}
\affiliation{Synergetic Innovation Center of Quantum Information and Quantum Physics, University of Science and Technology of China, CAS, Hefei, 230026, China}

\author{ G.-C. Guo}
\affiliation{CAS Key Laboratory of Quantum Information, University of Science and Technology of China, Hefei, 230026, China}
\affiliation{Synergetic Innovation Center of Quantum Information and Quantum Physics, University of Science and Technology of China, CAS, Hefei, 230026, China}


\date{\today}

\begin{abstract}
We introduce a scheme for remote entanglement generation for the photon polarization. The technique is based on transferring the initial frequency correlations to specific polarization-frequency correlations by local dephasing and their subsequent removal by frequency up-conversion. On fundamental level, our theoretical results show how to create and transfer entanglement, to particles which never interact, by means of local operations. This possibility stems from the  multi-path interference and its control in frequency space. For applications, the 
developed techniques and results allow for the remote generation of entanglement with distant parties without Bell state measurements and opens the perspective to probe frequency-frequency entanglement by measuring the polarization state of the photons.
\end{abstract}


\maketitle

\section{\label{I}Introduction}

The study of light frequency has had a key role in the development of modern physics. Even the first observations of quantum effects such as blackbody radiation and photoelectric effect relied on spectroscopic studies. Still nowadays frequency is one of the most utilized degrees of freedom (DOF) of light in photon based technologies. Indeed, frequency multiplexing of information is crucial for classical telecommunications and modern fluorescence imaging techniques and recently, a lot of effort has been put in exploiting the frequency DOF also for quantum-based information technologies \cite{zeilinger-2009, zhao_entangling_2014, cho_engineering_2014, roslund_wavelength-multiplexed_2014-1, Reimer1176, kobayashi_frequency-domain_2016}.

Photonic architectures are a natural candidate for realizing quantum networks since photons are fairly insensitive to environmental noise, easily manipulated, and efficient detection methods are readily available. However, the realization of future quantum networks requires reliable long-distance transmission of quantum information and entanglement. Although photons are the optimal quantum information carriers in long-distance quantum communication, the polarization DOF, which is most often utilized for processing the quantum information, is sensitive to noise in free space and optical fibres. Since other DOFs are less fragile to such noise, multi-DOF hyper-entangled quantum networks have been suggested as more reliable solution for quantum information transmission \cite{aolita_quantum_2007, xiao_efficient_2008, wang_quantum_2009, degreve-2012}.

\begin{figure}
\includegraphics[width= \linewidth]{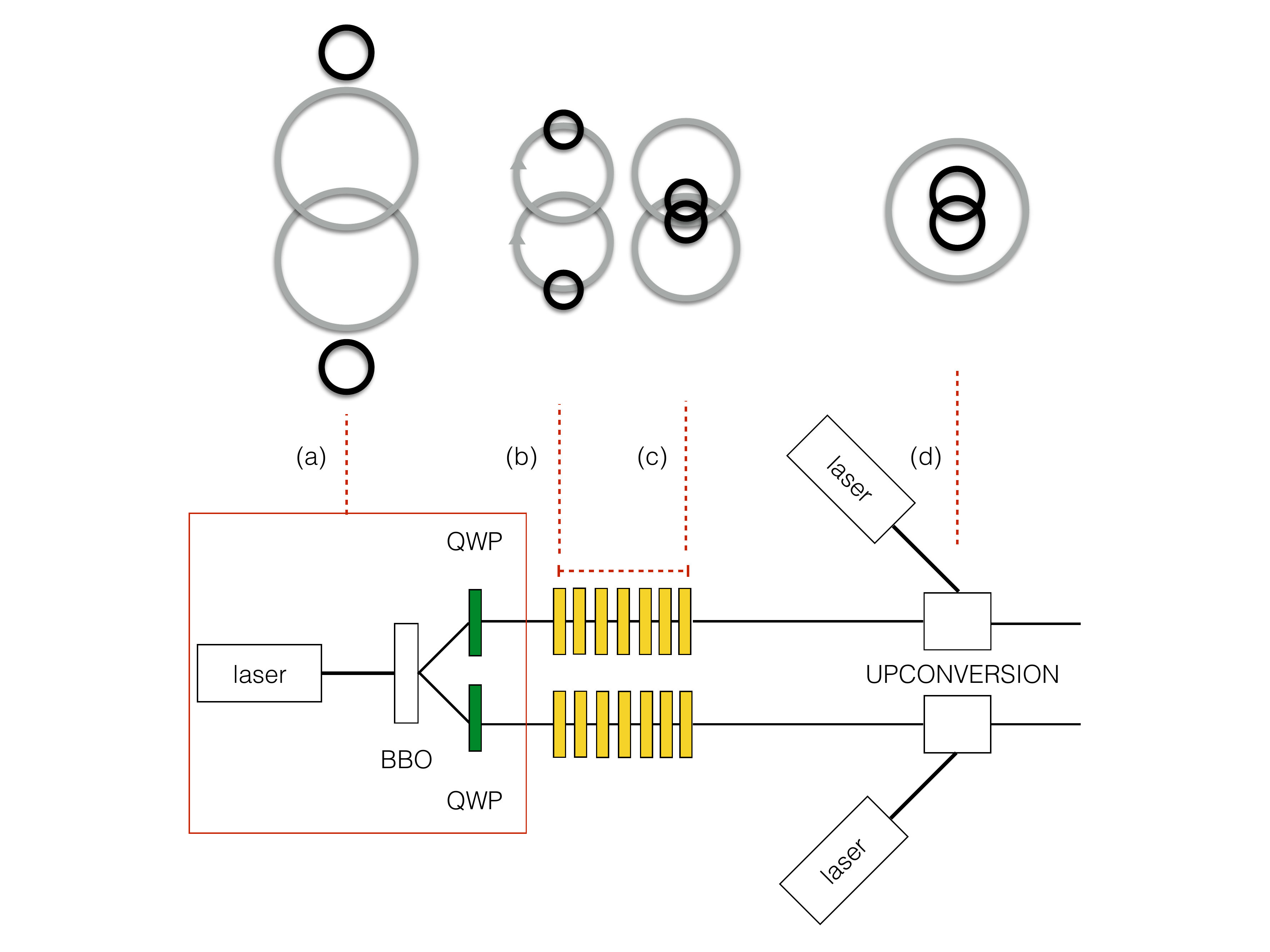}
\caption{\label{scheme}
A schematic picture of the protocol and corresponding optical set-up. (a) The state after parametric downconversion. The polarization state (black circles) is factorised and frequency state (grey circles) is entangled. See Eqs.~(1) and (2). (b) The polarization is coupled with the frequency in a quartz plate. See
 Eq.~(11). (c) The interaction time is fixed. See Eq.~(16). (d) Parametric upconversion erases the frequency information and produces entangled polarization state. See Eq.~(17).}
\end{figure}

For the operation of multi-DOF quantum networks transfering entanglement between the different DOFs is crucial \cite{nagali_quantum_2009}. In this paper we explore the possibility to transfer frequency entanglement to the polarization DOF after transmission of photons, thus allowing remote entanglement generation for the photon polarization. The protocol is performed in two steps: first the polarization DOF is coupled with the frequency in a local birefringent medium realizing local dephasing noise.  
Then, erasure of which frequency information, achievable via local frequency conversion \cite{takesue-2008, Tanzilli:2005aa, Ikuta:2011aa} is performed.  Besides the ideal case with discrete color (frequency bin) entanglement, we also explore the possibility of using continuous frequency entanglement which is naturally present in spontaneous parametric downconversion (SPDC) experiments as a consequence of energy conservation. Figure \ref{scheme} displays schematically the basic ideas and corresponding optical set-up.

The paper is structured as follows: Section \ref{II} descibes theoretically the optical setup for realizing the protocol. In section \ref{III} the ideal case with discrete color entanglement is presented and the connection with quantum erasure is discussed. Section \ref{IV} explores the possibility of using continuous frequency entanglement for the creation of polarization entanglement and section \ref{V} summarizes the results and discusses some future directions.

\section{\label{II} General formalism of the optical setup}

We consider a pair of photons subjected to local birefringent environments \cite{feng-2010, kwiat-2000}.
The same set up has also been used recently to study and implement nonlocal memory
effects \cite{Laine2012a,Laine2013b,Liu2013a}

The polarization degree of freedom of the photons constitutes the two-qubit open system whilst the continuous frequency degree of freedom of each photon forms the local environments.  
A common source of photon pairs is based on SPDC which allows to create and to control the  amount of initial entanglement between the frequencies of the two photons -- that is between the two environments in our scheme -- by controlling the properties of the down-conversion pump.  
The two photons created in SPDC travel along two arms, which we label by $a$ and $b$, and for the total system we 
have initially a polarization-frequency product state   
\begin{equation}\label{eq:in pol}
\ket{\Psi(0)}=\ket{\psi(0)}\otimes\int \int d\omega_a d\omega_b g(\omega_a,\omega_b)\ket{\omega_a,\omega_b}
\end{equation}
where $g(\omega_a,\omega_b)$ is the joint probability amplitude of finding a photon with frequency $\omega_a$ in arm $a$ and a photon with frequency $\omega_b$ in arm $b$ with the corresponding joint probability distribution $P(\omega_a,\omega_b)= |g(\omega_a,\omega_b)|^2$. Since one of our motivations is to develop a technique for remote creation of polarization entanglement, we choose as initial state a polarization product state
\begin{equation}
\label{Eq:Pini}
\ket{\psi(0)}=\frac{1}{2}\big(\ket{H_a}\ket{H_b}+\ket{H_a}\ket{V_b}+\ket{V_a}\ket{H_b}+\ket{V_a}\ket{V_b}\big)
\end{equation}
where $H$ ($V$) corresponds to horizontal (vertical) polarization.

The local system-environment -- or polarization-frequency -- interaction is obtained by inserting quartz plates along each arm. 
Due to birefringence in quartz plates, polarization and frequency of each photon interact with a local interaction Hamiltonian 
\begin{equation}\label{eq:deph ham}
\hat{H}_i=-\int d\omega_i \, \omega_i \left(n_H\ket{H_i}\bra{H_i}+n_V\ket{V_i}\bra{V_i}\right)\otimes \ket{\omega_i}\bra{\omega_i},
\end{equation}
where $i$ denotes one of the arms $a$ or $b$ and $n_{H(V)}$ is the index of refraction for horizontal (vertical) polarization component.

After an interaction time $\tau$ in each arm, the total system state reads
 \begin{eqnarray}\label{eq:deph state}
\ket{\Psi(\tau)}&&=\frac{1}{2}\int \int d\omega_a d \omega_b g(\omega_a,\omega_b)\Big(e^{i\tau n_H(\omega_a+\omega_b)}\ket{H_a}\ket{H_b} \nonumber
\\
&&+e^{i\tau n_V(\omega_a+\omega_b)}\ket{V_a}\ket{V_b}+e^{i\tau(n_H\omega_a+n_V \omega_b)}\ket{H_a}\ket{V_b}\nonumber
\\
&&+e^{i\tau(n_V\omega_a+n_H\omega_b)}\ket{V_a}\ket{H_b}\Big)\otimes\ket{\omega_a,\omega_b}.
\end{eqnarray} 
It is important to note here that the total system state, and the polarization state when tracing over the frequency, depend crucially on the joint initial frequency amplitude distribution $g(\omega_a,\omega_b)$. The two extreme cases correspond to i) having fully anticorrelated initial frequencies for the photons, i.e., $\omega_a+\omega_b=\omega_0$ where $\omega_0$ is the frequency of the pump ii) completely uncorrelated frequencies (wide SPDC pump). For the first case, it is easy to see from the r.h.s.~in the above equation that, when tracing over the frequency, the polarization components $\ket{H_a}\ket{H_b}$ and $\ket{V_a}\ket{V_b}$ retain well-defined and precise relative phase between them throughout the time evolution, since $\omega_a+\omega_b=\omega_0$, whereas the polarization components $\ket{H_a}\ket{V_b}$ and $\ket{V_a}\ket{H_b}$ dephase. For the second case, when there are no initial correlations between the frequencies of the photons, all polarization components contribute to dephasing. Obviously, in both of the cases during time evolution, the local interactions induce correlations between polarization and frequency.

Our aim is now to see whether we can, after the local dephasing interactions, create entanglement between the qubits by removing the generated system-environment correlations.  Indeed, we can use up-conversion to modify the frequency distributions of the photons, and at the same time remove the polarization-frequency correlations. However, a considerable challenge here is to design a scheme which allows the photon pairs having only polarization components $\ket{H_a}\ket{H_b}$ and $\ket{V_a}\ket{V_b}$ in Eq.(\ref{eq:deph state}) to be up-converted while not up-converting the components  $\ket{H_a}\ket{V_b}$ and $\ket{V_a}\ket{H_b}$. 

For this purpose, we consider local frequency up-conversion of each of the photons. For the calculations below, we assume that the shape of the pumps for each photon has the same structure and that the up-conversion process is local. Since we are interested in only those photons which do get up-converted, we do not use the full Hamiltonian to describe the process but deal with an operator corresponding to the matrix element of the Hamiltonian coupling the initial states into the up-converted states. Since we deal with single photons locally, the up-conversion process can be described with a local operator
 \begin{equation}\label{eq: up ham}
\hat{O}_i=\int d\nu_i \, P(\nu_i)\ket{\omega_i+\nu_i}\bra{\omega_i}.
\end{equation}
Here, $i$ takes values $a$ or $b$ corresponding to the two arms and the probability distribution $P(\nu_i)$ for the frequency of the laser pumps also describes the coupling strength or efficiency of the up-conversion process. 
This operator maps a given frequency $\omega_i$ of the photon in arm $i$ to frequency $\omega_i+\nu_i$. 
If we assume that the average of the local single photon frequency distribution $P(\omega_i)$ before the up-conversion and the average of the up-conversion pump distribution  $P(\nu_i)$ are equal, but the width of the latter is much larger the former, then after the up-conversion the local frequency distributions of the photons are identical. Moreover, since the frequency of the up-converted photon is random - within the limits of used pump shapes and corresponding distributions, we have also removed all the possible frequency correlations between the photons in addition of removing the polarization-frequency correlations. 

To make the description more rigorous and quantitative, let us consider now in detail the influence of the up-conversion process with operator (\ref{eq: up ham}) on the total state (\ref{eq:deph state}).
Applying operator $\hat{O}$ on both arms, the system-environment state after local up-conversion in the two arms reads
\begin{widetext}
\begin{eqnarray}\label{eq:up state}
\ket{\Psi(\tau)}_u= \frac{1}{2}\int d\omega_a\int d\omega_b \int  d \nu_a\int 
 d \nu_b \, &&g(\omega_a, \omega_b)P(\nu_a)P(\nu_b)\times \Big(e^{i\tau n_H(\omega_a+\omega_b)}\ket{H_a}\ket{H_b}+e^{i\tau(n_H\omega_a+n_V \omega_b)}\ket{H_a}\ket{V_b}
\nonumber
\\
&&+e^{i\tau(n_V\omega_a+n_H\omega_b)}\ket{V_a}\ket{H_b}+e^{i\tau n_V(\omega_a+\omega_b)}\ket{V_a}\ket{V_b}\Big)
\otimes\ket{\omega_a+\nu_a ,\omega_b+\nu_b}.
\end{eqnarray}
\end{widetext}
Let us for the sake of convenience use a more compact notation in the rest of the paper by defining $\ket{H_a}\ket{V_b}\equiv \ket{HV}$. 
We can obtain the reduced state of the open system (polarization) by tracing over the frequency (environment).
Formally, the elements of the polarization state density matrix can be therefore expressed as
\begin{widetext}
\begin{equation}\label{eq:up matrix}
\bra{\lambda \mu}\rho(\tau)\ket{\lambda^{'} \mu^{'}}=\frac{1}{4}\int d\omega_a \int d\omega^{'}_a \int d \omega_b \int d \omega^{'}_b \, g(\omega_a,\omega_b)g^{*}(\omega^{'}_a,\omega^{'}_b) \mathrm{exp}[ i\tau(n_{\lambda}\omega_a+n_{\mu}\omega_b-n_{\lambda^{'}}\omega^{'}_a-n_{\mu^{'}}\omega^{'}_b)] E(\omega_a-\omega^{'}_a,\omega_b-\omega^{'}_b),
\end{equation}
\end{widetext}
where $\lambda, \mu, \lambda^{'}, \mu^{'}=H,V$ and the function the function $E$ is defined as
\begin{eqnarray}
E(\omega_a-\omega^{'}_a,\omega_b-\omega^{'}_b)=&&\int d\nu_a P(\nu_a)P(\nu_a+\omega_a-\omega^{'}_a) 
\\
&&\times\int d\nu_b P(\nu_b)P(\nu_b+\omega_b-\omega^{'}_b).\nonumber
\end{eqnarray}
It is evident from Eq.~\eqref{eq:up matrix} that the final polarization state depends critically on the function  $E$ whose form is in turn crucially connected to the efficiency of erasing prior information on the frequency of the photons  and polarization-frequency correlations.
If we assume a Gaussian frequency distribution for the up-conversion pump, such that $P(\nu)=\mathrm{exp}[\frac{-(\nu-\nu_0)^2}{2\sigma^2}]/\sigma\sqrt{2\pi}$, where $\sigma$ and $\nu_0$ are the standard deviation and the average, respectively,  the function $E$ takes the form
\begin{equation}\label{eq: erasure}
E(\omega_a-\omega^{'}_a,\omega_b-\omega^{'}_b)=\frac{\mathrm{e}^\frac{{-(\omega_a-\omega^{'}_a)^2}}{4\sigma^2}}{2\sigma \sqrt{\pi}}\times \frac{\mathrm{e}^\frac{{-(\omega_b-\omega^{'}_b)^2}}{4\sigma^2}}{2\sigma \sqrt{\pi}}.
\end{equation} 
 We have now two extreme cases depending on whether $\sigma \gg \omega_i-\omega^{'}_i$ or  $\sigma \ll \omega_i-\omega^{'}_i$.
 In the former case, with very wide up-conversion pump, $E$ can be approximated by 
 \begin{equation}
 \label{eq:eappro}
 E \approx 1/4\pi \sigma^2.
 \end{equation} 
 In the opposite case, with narrow pump, we have $E(\omega_a-\omega^{'}_a,\omega_b-\omega^{'}_b)\approx \delta(\omega_a-\omega^{'}_a)\delta(\omega_b-\omega^{'}_b)$.
 
We have now presented the general expression for the polarization state of the pair of photons after the local dephasing processes followed by up-conversion. Before introducing the results and discussing under which conditions and how much entanglement can be generated remotely, we will briefly present an idealized case. This illustrates the basic mechanism and provides intuition on the reason why entanglement generation is possible in our scheme. 

\section{\label{III}Ideal erasure with discrete color entanglement}

As an ideal case we consider frequency states where, instead of the finite width frequency distributions, the state is in a discrete color entangled state
$1/\sqrt{2}(\ket{\omega_1,\omega_2}+\ket{\omega_2,\omega_1})$. The preparation of such states has been studied in e.g. \cite{zeilinger-2009}. With initial polarization product state \eqref{Eq:Pini} and after the local dephasing interaction given by the Hamiltonian \eqref{eq:deph ham}
the total system state is
\begin{eqnarray}\label{eq: deph state ideal}
\ket{\Psi(\tau)}=&&\frac{\mathrm{e}^{i\tau n_H\omega_0}}{2}\Big(\ket{\psi_0(\tau)}\otimes(\ket{\omega_1,\omega_2}+\ket{\omega_2,\omega_1})
\\
&&+\ket{\psi_1(\tau)}\otimes \ket{\omega_1,\omega_2}+\ket{\psi_2(\tau)}\otimes \ket{\omega_2,\omega_1}\Big),\nonumber
\end{eqnarray}
where 
\begin{eqnarray}\label{eq: subspaces}
&&\ket{\psi_0(\tau)}=\frac{1}{\sqrt{2}}\left( \ket{HH}+\mathrm{e}^{i\tau \Delta n \omega_0}\ket{VV}\right),
\\
&&\ket{\psi_1(\tau)}=\frac{1}{\sqrt{2}}\left(\mathrm{e}^{i\tau \Delta n \omega_2}\ket{HV}+\mathrm{e}^{i\tau \Delta n \omega_1}\ket{VH}\right),
\\
&&\ket{\psi_2(\tau)}=\frac{1}{\sqrt{2}}\left(\mathrm{e}^{i\tau \Delta n \omega_1}\ket{HV}+\mathrm{e}^{i\tau \Delta n \omega_2}\ket{VH}\right).
\end{eqnarray}
Here $\Delta n=n_V-n_H$ and $\omega_0=\omega_1+\omega_2$. Equation \eqref{eq: deph state ideal} shows that the polarization subspace spanned by $\ket{\psi_0(\tau)}$ ($\ket{HH}$, $\ket{VV}$) always remains factorized from the environment. On contrast, the behaviour of the subspace spanned $\ket{\psi_{1,2}(\tau)}$ ($\ket{HV}$, $\ket{VV}$) is different.
To realize this fact, consider the inner product $\langle \psi_1(\tau)\ket{\psi_2(\tau)}=\cos(t\Delta n \Delta \Omega)$, where we have defined $\Delta \Omega= \omega_1-\omega_2$. Hence we have periodically times $\tau=\tau_c$ at which $\ket{\psi_1(\tau_c)}=\ket{\psi_2(\tau_c)}$ and $\tau=\tau_d$ where $\ket{\psi_1(\tau_d)}=-\ket{\psi_2(\tau_d)}$. 
Therefore, the total system states at these two points of time are
 \begin{eqnarray}\label{eq: state ideal cest}
\ket{\Psi(\tau_c)}=&&\frac{\mathrm{e}^{i\tau_c n_H\omega_0}}{2}\Big[\left( \ket{\psi_0(\tau_c)}+ \ket{\psi_1(\tau_c)}\right) \nonumber
\\
&&\otimes(\ket{\omega_1,\omega_2}+\ket{\omega_2,\omega_1}) \Big]
\end{eqnarray}
and
\begin{eqnarray}\label{eq: state ideal dest}
\ket{\Psi(\tau_d)}=&&\frac{\mathrm{e}^{i\tau_d n_H\omega_0}}{2}\Big[\ket{\psi_0(\tau_d)}\otimes(\ket{\omega_1,\omega_2}+\ket{\omega_2,\omega_1}) \nonumber
\\
&&+\ket{\psi_1(\tau_d)}\otimes (\ket{\omega_1,\omega_2} -\ket{\omega_2,\omega_1})\Big].
\end{eqnarray}
Obviously, the former is a product polarization-frequency state. For the latter, the initial frequency entanglement has been transferred to full polarization-frequency entanglement. Note also that here the measurement of polarization would reveal the full information about the frequency state. In order to create polarization entanglement, we want to maintain only the component including the state $\ket{\psi_0}$ in Eq.~\eqref{eq: state ideal dest}. To this end, consider an ideal upconversion in which we map locally any initial frequency of each of the photons to a fixed frequency 
$\omega_u$. Therefore, all  frequency vectors will be mapped to $\ket{\omega_u,\omega_u}$ and the total state of the pair changes to
\begin{eqnarray}\label{eq: EG pol}
\ket{\Psi(\tau_d)}&=&\mathrm{e}^{i\tau_d n_H\omega_0}\left(\ket{\psi_0(\tau_d)}+\ket{\psi_1(\tau_d)}-\ket{\psi_1(\tau_d)}\right) \nonumber 
\\
&&\otimes\ket{\omega_u,\omega_u} \nonumber \\
&& = \mathrm{e}^{i\tau_d n_H\omega_0}\ket{\psi_0(\tau_d)}\otimes\ket{\omega_u,\omega_u}.
\end{eqnarray}
which includes fully entangled polarization state $\ket{\psi_0(\tau_d)}$ for the open system.

The procedure depicted above can be seen as an erasure procedure with entangled tags \cite{scully-1982}. With discrete color entangled frequency state, the frequency acts as a which way tag for the polarization and the entanglement transfer process is similar to that described in \cite{PhysRevA.60.827}. However here, we have instead of one which way tag, two entangled tags that locally interact with the system. Such setup allows, instead of disentanglement erasure, a transfer of entanglement to an initially uncorrelated system.

\begin{figure}
\includegraphics[width= \linewidth]{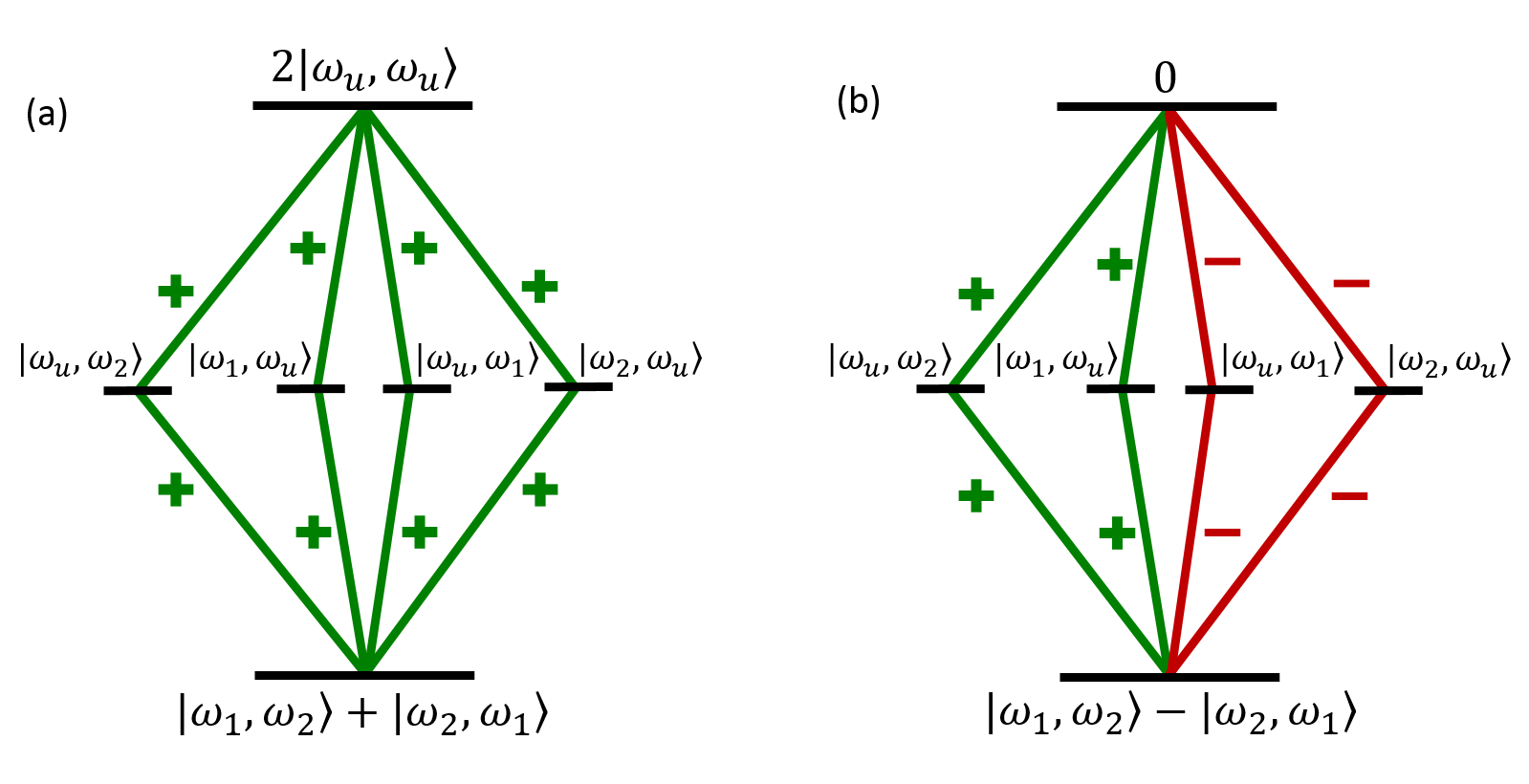}
\caption{\label{scheme2}Schematic description of interference of mode paths for local up-conversion of both photons. 
(a) Frequency component $\ket{\omega_1,\omega_2}+\ket{\omega_2,\omega_1}$, which is associated to polarization state $|\psi_0\rangle$ in Eq.~(\ref{eq: state ideal dest}). 
(b) Frequency component $\ket{\omega_1,\omega_2}-\ket{\omega_2,\omega_1}$ which is associated to polarization state $|\psi_1\rangle$ in Eq.~(\ref{eq: state ideal dest}).
After up-conversion of only one of the photons, initial vectors  have four mode paths in the middle of the figure with corresponding signs due initial relative phases. After the up-conversion of the other photon, mode paths interfere constructively in $(a)$ and destructively in (b).}
\end{figure}

This process of entanglement generation is also linked to interference of mode paths during the erasure process. As we depict in Fig.~\ref{scheme2}, and according to Eq.~\eqref{eq: state ideal dest}, when we upconvert frequency of each photon to $\omega_u$, components including two orthogonal vectors $\ket{\omega_1,\omega_2}+\ket{\omega_2,\omega_1}$ and $\ket{\omega_1,\omega_2}-\ket{\omega_2,\omega_1}$ interfere differently. While the first one leads to constructive interference along the mode paths, the latter comes with destructive interference of them, thus eliminating  $\ket{\psi_1(\tau)}$ and allowing the generation of entanglement in the open system. 

The simplifications in the above scheme are two-fold even in the case of initially fully anticorrelated frequencies when $\omega_1 + \omega_2=\omega_0$. Firstly, in realistic case one has large number of $\omega_1$ and $\omega_2$ pairs, whose sum is equal to $\omega_0$, since the frequency distributions of the created photons in SPDC have finite but non-zero widths. Secondly, the upcoversion does not produce single frequency photons with $\ket{\omega_u}$ but they have wide and continuous distribution. Therefore, efficient entanglement generation with a more realistic model requires total constructive and destructive interference paths consisting of all  possible
$\omega_1$ and $\omega_2$ pairs for all possible $\omega_u$'s.  Despite of this subtle issue, we show in the next section that efficient entanglement generation is possible also with continuosly entangled photon pairs.

\section{\label{IV} Continous frequency entanglement}

In the previous section we showed how discrete color entanglement can be transferred to the polarization DOF via an erasure procedure. Now, we want to explore, whether continuous frequency entanglement would allow for inter-DOF transfer of entanglement. Such entanglement occurs naturally between two photons after a downconversion process due to energy conservation.

To this end, let us study how the form of the initial frequency distribution, e.g. its width, and the degree of initial frequency correlations influence the amount of created polarization entanglement.  
Moreover, we consider single and double peak Gaussian distributions -- that can be associated with Markovian and non-Markovian dynamics respectively \cite{Liu:2011aa}.

\subsection{\label{IV double}Double-peak initial frequency distribution}

To have a double-peak distribution, we consider a symmetric sum of two bivariate Gaussian distributions with the same variance and correlation coefficient but different mean values. The first bivariate distribution is peaked at
 $(\Omega_1,\Omega_2)$, while the other has its mean value at $(\Omega_2,\Omega_1)$. 
We also have  $\Delta\Omega = \Omega_2-\Omega_1$  and $\Omega_2+\Omega_1=\omega_0$ where $\omega_0$ is the central frequency of the SPDC laser pump. Therefore, the frequency distribution of the initial state of the environment is (apart from the normalization factor)
\begin{equation}\label{eq: double peak dis}
\vert g(\omega_a,\omega_b)\vert^2= P_1(\omega_a,\omega_b)+P_2(\omega_a,\omega_b),
\end{equation}
where we have defined 
\begin{equation}\label{eq: double peak prob}
P_{1(2)}(\omega_a,\omega_b)=\frac{1}{2\pi \sqrt{\mathrm{det} C}}e^{-\frac{\big(\vec{\omega}- \langle\vec{\omega_{1(2)}}\rangle\big) ^{T}C^{-1}\big(\vec{\omega}- \langle\vec{\omega_{1(2)}}\rangle\big)}{2}}.
\end{equation}
Here, $\vec{\omega}=(\omega_a,\omega_b)^T$, $\langle\vec{\omega_1}\rangle=(\Omega_1,\Omega_2)^T$, $\langle\vec{\omega_2}\rangle=(\Omega_2,\Omega_1)^T$ and $C$ is the covariance matrix
\begin{equation}\label{eq: C matrix}
C=\left (
\begin{array}{cc}
\delta^2 & k \delta^2 \\ 
k \delta^2 & \delta^2
\end{array} 
\right ),
\end{equation}
where $k$ is the correlation coefficient, and $\delta$ is the standard deviation. 

To have the probability amplitudes $g(\omega_a,\omega_b)$, we take the square root of the probability distribution defined in Eq.~\eqref{eq: double peak dis}. This is because the phases of the frequency amplitudes are 
non-random~\cite{Ou99} and we take the initial amplitudes to be real.
 Therefore, the initial joint frequency probability amplitudes of the photon pair are given by 
\begin{equation}\label{eq: probability amp double}
g(\omega_a,\omega_b)=\sqrt{P_1(\omega_a,\omega_b)+P_2(\omega_a,\omega_b)}.
\end{equation}
We consider the case where the peaks of the two Gaussians given by $P_1(\Omega_1,\Omega_2)$ and $P_2(\Omega_2,\Omega_1)$  are well-separated, $\Delta \Omega \gg \delta$, and therefore we can make the following approximation
\begin{equation}\label{eq: probability amp sep peaks}
g(\omega_a,\omega_b)\approx \sqrt{P_1(\omega_a,\omega_b)}+\sqrt{P_2(\omega_a,\omega_b)}.
\end{equation}
Substituting amplitudes \eqref{eq: probability amp sep peaks} into Eq.~\eqref{eq:up matrix}, and considering a perfect erasure procedure with wide up-conversion pump [c.f.~ Eq.~\eqref{eq:eappro}],
yield the following elements for the reduced polarization state density matrix (dropping all common factors and apart from normalization since we are interested only in the up-converted part)
\begin{widetext}
\begin{eqnarray}\label{eq: double peak elements}
&&\langle HH \vert \rho(\tau) \ket{HH}=2 e^{-4 (1+k) n_H^2 \tau^2 \delta ^2},\label{eq: double peak elements 1}
\\
&&\langle HH \vert \rho(\tau) \ket{HV}= e^{- \left((3+2 k) n_H{}^2 +2 k n_H n_V+n_V {}^2\right)\tau^2 \delta ^2} \left(e^{-i \tau \text{$\Delta $n} \text{$\Omega $1}}+e^{-i \tau \text{$\Delta $n} \text{$\Omega $2}}\right),\label{eq: double peak elements 2}
\\
&&\langle HH \vert \rho(\tau) \ket{VH}=e^{- \left((3+2 k) n_H{}^2 +2 k n_H n_V+n_V {}^2\right)\tau^2 \delta ^2} \left(e^{-i \tau \text{$\Delta $n} \text{$\Omega $1}}+e^{-i \tau \text{$\Delta $n} \text{$\Omega $2}}\right),\label{eq: double peak elements 3}
\\
&&\langle HH \vert \rho(\tau) \ket{VV}=2 e^{-2 (1+k) \left(n_H^2+n_V^2\right) \tau^2 \delta ^2-i \tau \text{$\Delta $n} \omega _0},\label{eq: double peak elements 4}
\\
&&\langle HV \vert \rho(\tau) \ket{HV}=e^{-2 \left(n_H{}^2+2 k n_H n_V+n_V{}^2\right) \tau^2 \delta ^2}\Big(1+\cos(\tau \Delta n  \Delta \Omega )\Big),\label{eq: double peak elements 5}
\\
&&\langle HV \vert \rho(\tau) \ket{VH}=e^{-2 \left(n_H{}^2+2 k n_H n_V+n_V{}^2\right) \tau^2 \delta ^2}\Big(1+\cos(\tau \Delta n  \Delta \Omega )\Big),\label{eq: double peak elements 6}
\\
&&\langle HV \vert \rho(\tau) \ket{VV}=e^{- \left((3+2 k) n_V{}^2 +2 k n_H n_V+n_H {}^2\right)\tau^2 \delta ^2} \left(e^{-i \tau \text{$\Delta $n} \text{$\Omega $1}}+e^{-i \tau \text{$\Delta $n} \text{$\Omega $2}}\right),\label{eq: double peak elements 7}
\\
&&\langle VH \vert \rho(\tau) \ket{VH}=e^{-2 \left(n_H{}^2+2 k n_H n_V+n_V{}^2\right) \tau^2 \delta ^2}\Big(1+\cos(\tau\Delta n  \Delta \Omega )\Big),\label{eq: double peak elements 8}
\\
&&\langle VH \vert \rho(\tau) \ket{VV}=e^{- \left((3+2 k) n_V{}^2 +2 k n_H n_V+n_H {}^2\right)\tau^2 \delta ^2} \left(e^{-i \tau \text{$\Delta $n} \text{$\Omega $1}}+e^{-i \tau \text{$\Delta $n} \text{$\Omega $2}}\right),\label{eq: double peak elements 9}
\\
&&\langle VV \vert \rho(\tau) \ket{VV}=2 e^{-4 (1+k) n_V^2 \tau^2 \delta ^2}.\label{eq: double peak elements 10}
\end{eqnarray}
\end{widetext}
By setting $\delta=0$ and $k=-1$, corresponding to sharp local frequency peaks and fully entangled initial state, and having $\tau=\tau_d$ indicating that $\cos(\tau_d \Delta n  \Delta \Omega)=-1$, it is quite straigthfoward to see that we recover the ideal case result presented in the previous section [c.f.~Eqs.~\eqref{eq: subspaces} and \eqref{eq: EG pol}]. In general, it is worth noting that the solutions above contain oscillatory terms with $\cos(\tau \Delta n  \Delta \Omega)$ and exponential damping terms such as e.g. $\exp{\left[-2 \left(n_H{}^2+2 k n_H n_V+n_V{}^2\right) \tau^2 \delta ^2\right]}$ for the matrix element $\langle HV \vert \rho(\tau) \ket{HV}$ [c.f.~Eq.~\eqref{eq: double peak elements 5}] - one of the terms we want to minimize in order to maximize the entanglement generation with $\ket{HH}$ and  $\ket{VV}$. 

This clearly shows that we have two different time scales governing the entanglement generation: one given by the frequency difference between the peaks $\Delta \Omega$ and the other one related to the frequency width $\delta$. 
In particular, for fully anticorrelatated initial frequencies of the photons ($k=-1$), we have 
e.g.~$\exp{\left[-2 \left(n_H{}^2+2 k n_H n_V+n_V{}^2\right) \tau^2 \delta ^2\right]}=\exp{\left[-2 \Delta n^2\tau^2 \delta ^2\right]}$. Thereby, entanglement generation is possible both in the short interaction time non-Markovian region controlled by $\Delta \Omega$ and in the asymptotic long-time limit controlled by $\delta$.

\begin{figure}[t]
\includegraphics[width=\linewidth]{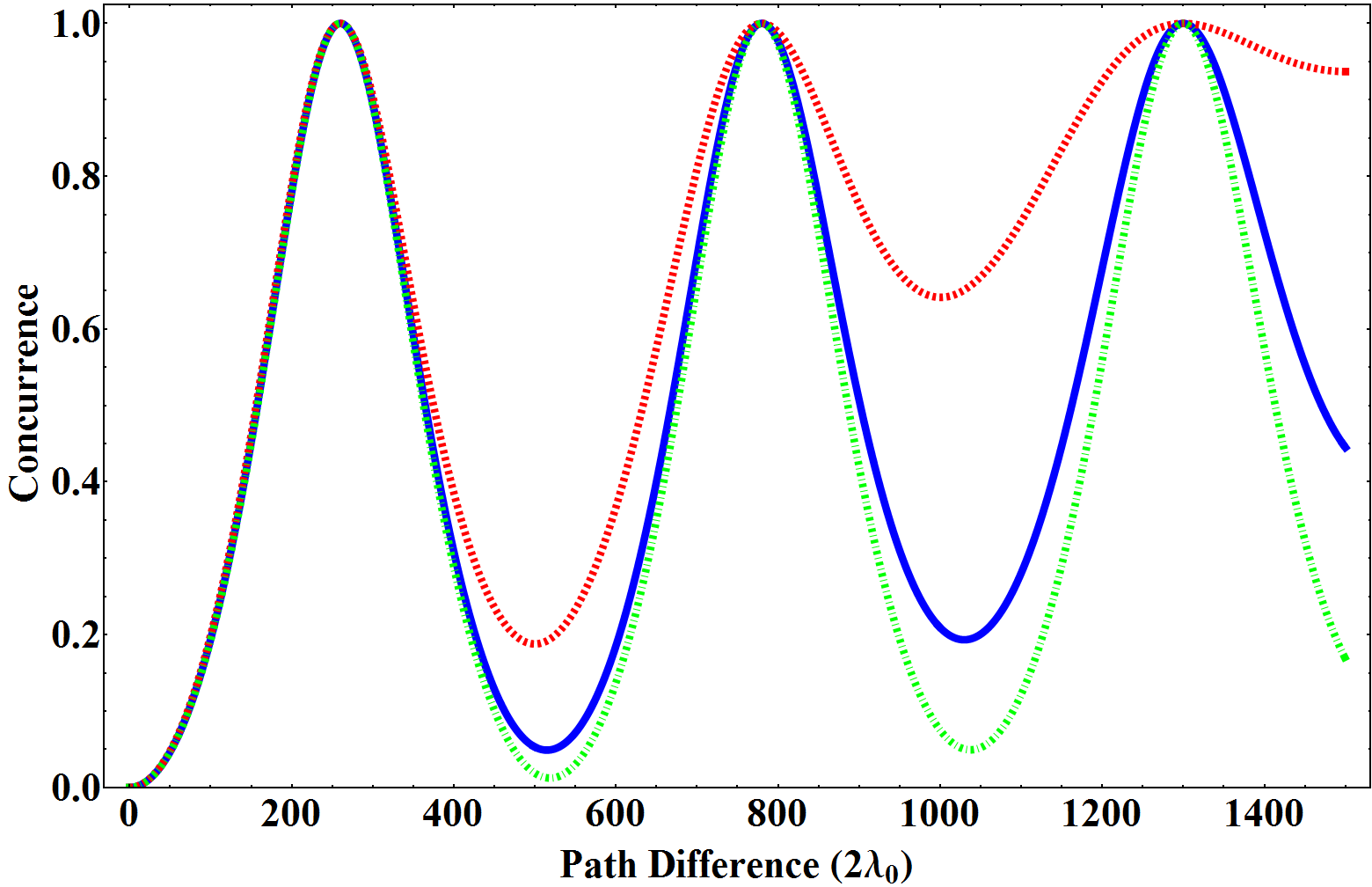}
\caption{\label{fig3}Polarization entanglement as a function of path difference for double-peak case and changing the widths of the frequency distribution. Here $\lambda_0=780$ \textit{nm}, $k=-1$, and $\Delta \Omega =3 \textit{nm}$; dot-dashed green line when the width (FWHM) of the local peaks is $0.125$ \textit{nm}, solid blue line when local the widths are $0.25$ \textit{nm}, and dashed red line when the local widths are $0.5$ \textit{nm}.}
\end{figure}

This observation is confirmed by the results presented in Fig.~\ref{fig3} where we plot the entanglement quantified by concurrence \cite{concurrence} as a function of the dephasing interaction time for $k=-1$.  Here, the interaction time is measured by the path difference
$\Delta n \times L/2\lambda_0$ with $\lambda_0=2\pi c/\omega_0$, and one can use $\tau=L/c$ to convert back to real time.
For very narrow initial frequency distributions, close to the case presented in the previous sections, we have periodic creation of large amount of entanglement (green curve). Increasing the width of the initial frequency distribution allows one to reach high-values of entanglement faster in the asymptotic regime (blue and red curves). Thereby one has two viable options: to control very precisely the interaction time in the short time region in order to exploit the peaks of the oscillation, or to use longer interaction times where the condition on the precise control of the interaction time can be relaxed.

\begin{figure}[t]
\includegraphics[width=\linewidth]{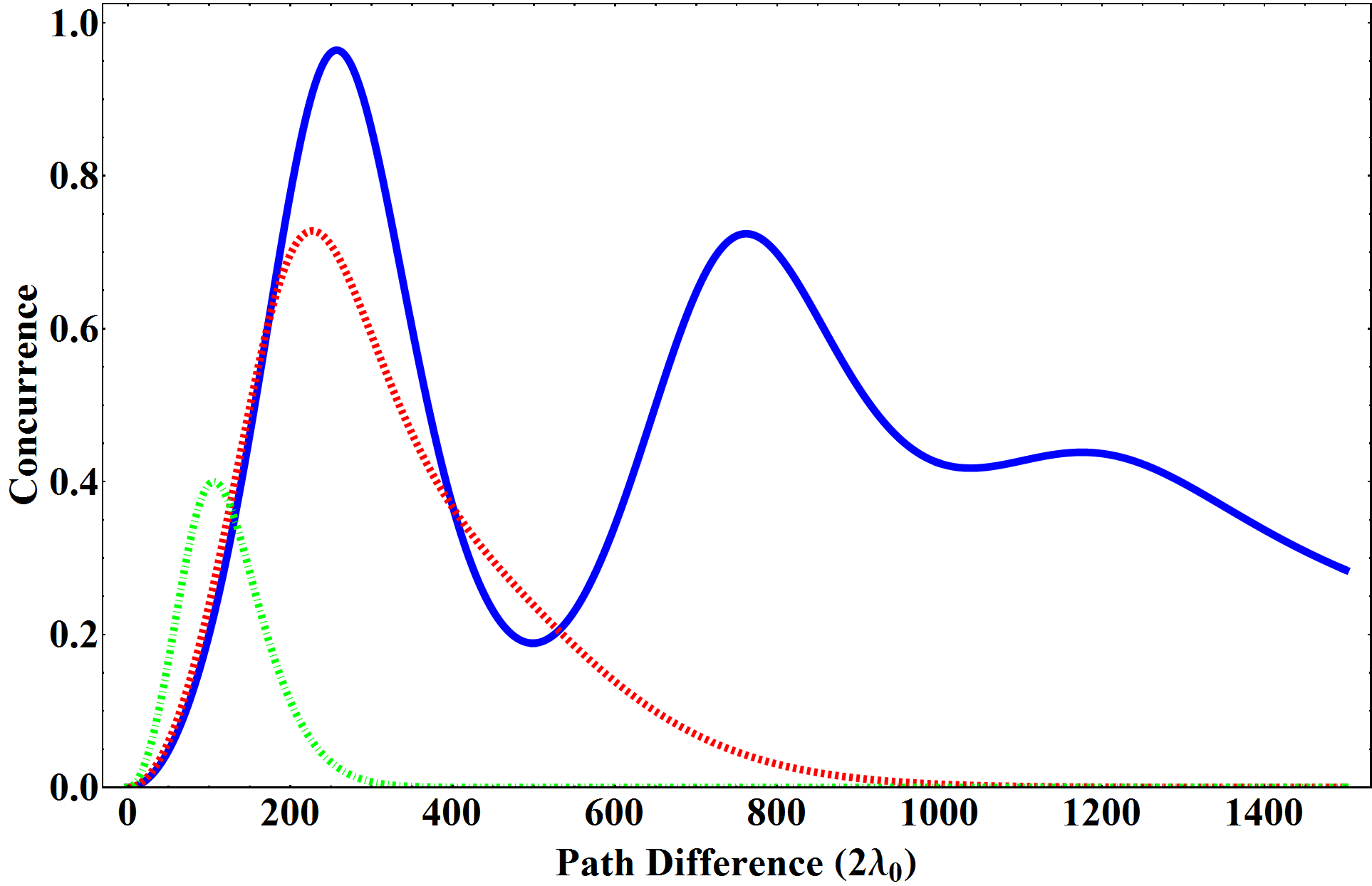}
\caption{\label{fig4}Polarization entanglement as a function of path difference for double-peak case and reducing the amount of initial correlations. Here $\lambda_0=780$ \textit{nm}, $n_H= 1.51004$, $n_V=1.54360$,  separation between two local peaks is 3 \textit{nm}, and local widths (FWHM) are $0.5$ \textit{nm}; solid blue line when $k=-0.999$, dashed red line  when $k=-0.99$, and dot-dashed green line for $k=-0.9$.}
\end{figure}

Reaching full frequency anticorrelations $k=-1$ may not always be easy, and therefore we are also interested in how reducing the amount of correlations influences the efficiency of the entanglement generation. Note, however that $k=-1.0$ and $k=-0.99995$ were recently used in an experiment exploiting nonlocal memory effects for efficient superdense coding~\cite{Liu2016a}.
Figure \ref{fig4} shows the results for $k>-1$. We see that for long interaction times the amount of generated entanglement decreases and ultimately tends to zero. However, the short time oscillations are still present and can be used to generate entanglement even though the maximum value of entanglement reached in the first oscillation peak diminishes for lower and lower initial frequency correlations.
Obviously, not having perfect frequency correlations means that the delicate balance required for interference in up-conversion for full entanglement generation is disturbed and the longer is the dephasing  time, the more prominent is the influence of imperfect initial correlations.  

We have already explained how -- in the double-peak case -- the entanglement generation is characterized by short time periodic and long-time asymptotic behaviour. The presence of the latter feature opens the possibility to generate entanglement also with initial single-peak frequency distribution and without the appaerance of non-Markovian memory effects. We consider this scenario in the next subsection.

\subsection{Single-peak initial frequency distribution}

We now consider a bivariate Gaussian distribution to describe the initial joint frequency probability distribution  of the photon pair
\begin{equation}\label{eq: single peak prob}
\vert g(\omega_a,\omega_b)\vert^2= \frac{1}{2\pi \sqrt{\mathrm{det} C}}e^{-\frac{\big(\vec{\omega}- \langle\vec{\omega}\rangle\big) ^{T}C^{-1}\big(\vec{\omega}- \langle\vec{\omega}\rangle\big)}{2}}.
\end{equation}
Here, $\vec{\omega}$ and $C$ are similar to the double peak case, but one has $\langle\vec{\omega}\rangle=(\omega_0/2,\omega_0/2)^T$, i.e. the Gaussian has mean value at  $(\omega_0/2,\omega_0/2)$.
As before, the probability amplitudes are given by the square root of the probability distribution defined in \eqref{eq: single peak prob}. 
We can again use the solution from Eqs. (\ref{eq: double peak elements 1}-\ref{eq: double peak elements 10}) by setting $\Omega_1=\Omega_2=\omega_0/2$ and $\Delta \Omega=0$. 
Obviously, the oscillatory parts in the solutions vanish since $\cos(\tau \Delta n  \Delta \Omega) = 0$ when $\Delta \Omega=0$
and we are left with exponential damping terms for the long-time region and for density matrix elements to be minimized in order to maximize the entanglement generation.

\begin{figure}[t]
\includegraphics[width=\linewidth]{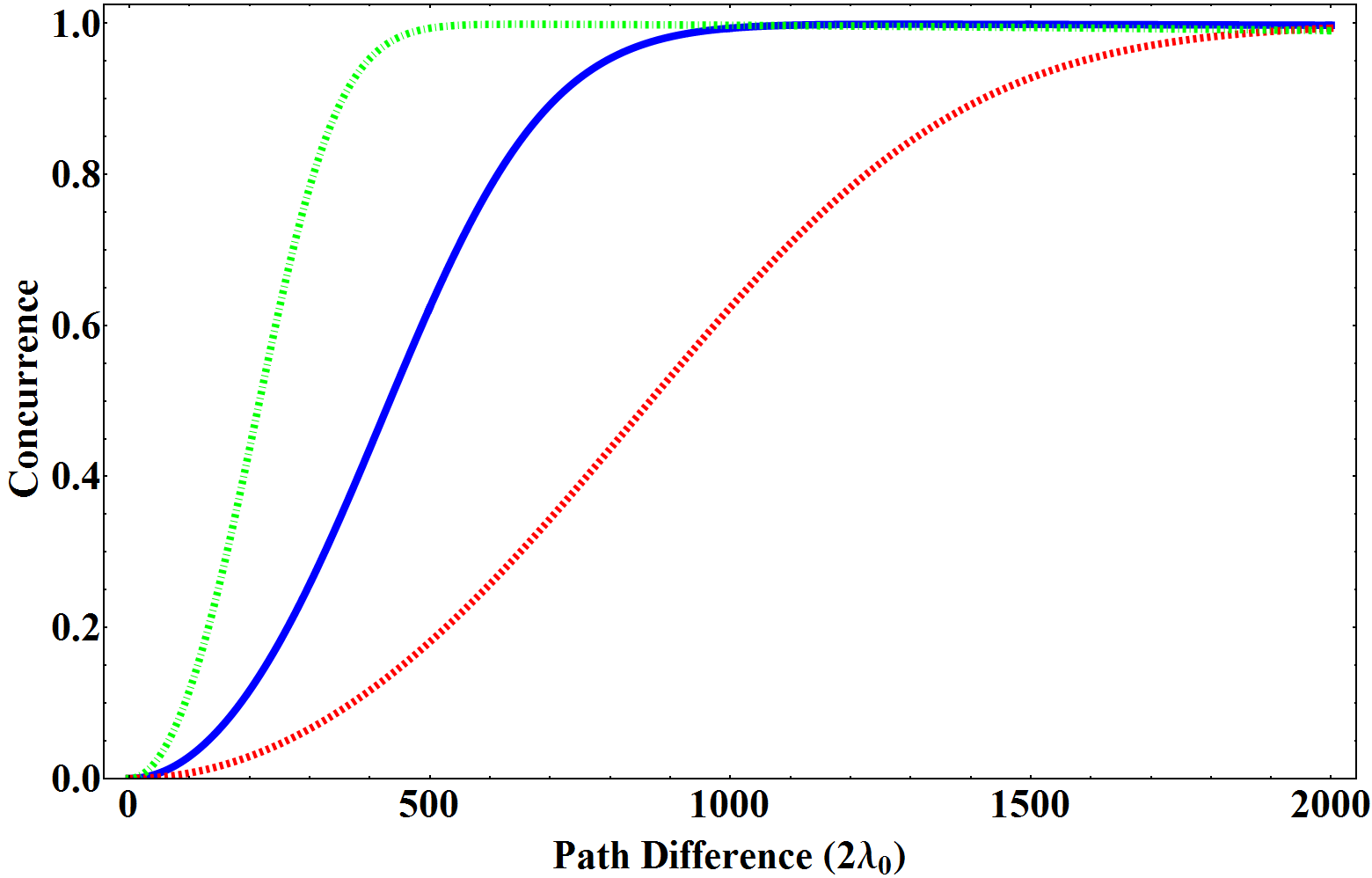}
\caption{\label{fig5} Polarization entanglement of single-peak case as a function of path difference for different widths of the initial frequency distribution with $k=-1$. Here $\lambda_0=780$ \textit{nm}.Dot-dashed green line when the local width (FWHM) is $2$ \textit{nm}, solid blue line when the  local width is $1$ \textit{nm}, and dashed red line when the local width is $0.5$ \textit{nm}.}
\end{figure}

Indeed, the results in Fig.~\ref{fig5} for $k=-1$ show the monotonic increase in the amount of generated entanglement as function of the dephasing time. The three curves correspond to different initial widths $\delta$ of the frequency distributions. The larger is the initial width, the faster is the entanglement generation. In other words, the dephasing for unwanted polarization components preventing their up-conversion is faster the wider is the frequency window of the environment.   

\begin{figure}[t]
\includegraphics[width=\linewidth]{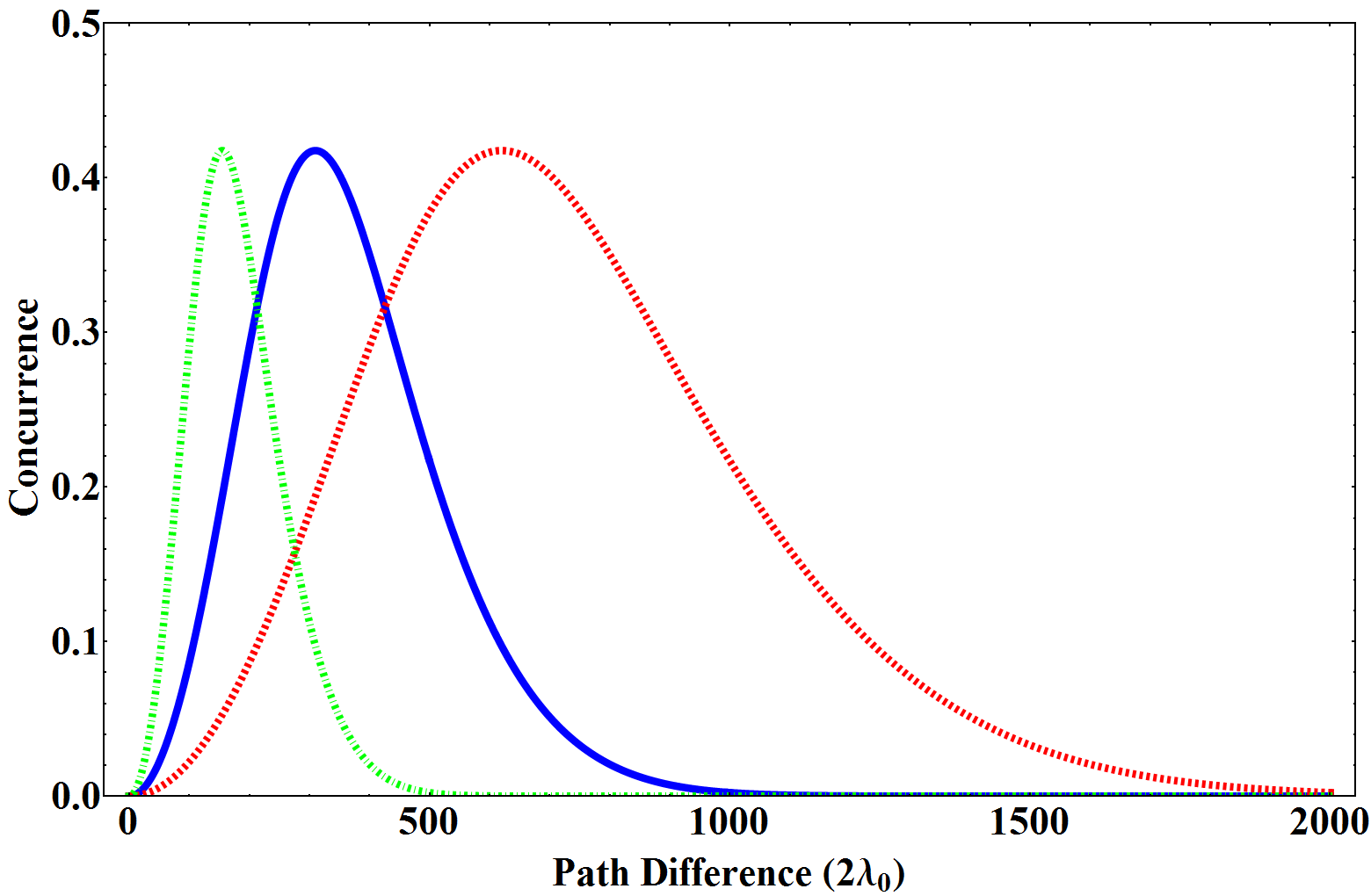}
\caption{\label{fig6}Polarization entanglement of single-peak case as a function of path difference for different widths of the initial frequency distribution with $k=-0.999$. Here $\lambda_0=780$ \textit{nm}, $n_H= 1.51004$, $n_V=1.54360$, dot-dashed green line when the local width (FWHM) is $2$ \textit{nm}, solid blue line when the local width is $1$ \textit{nm}, and dashed red line  when the  local width is $0.5$ \textit{nm}.}
\end{figure}

We saw before for the double-peak case that the entanglement generation is quite sensitive when the amount of initial frequency correlations are reduced. Figure~\ref{fig6} shows similar results as in Fig.\ref{fig5} but for $k=-0.999$. This result demonstrates that even though one can still reach considerable amount of entanglement, the process is more sensitive to imperfections in correlations compared to double-peak case. The results also show that one can make the required interaction time shorter by widening the initial frequency distribution but the maximun value of entanglement remains the same for a given value of the correlation coefficient $k$.

\section{\label{V}Discussion and conclusions}

We have shown how to convert initial frequency entanglement between two photons to their polarization entanglement. One of the interesting features here is that this can be achieved by local interactions and operations only without the need for any ancillary photons. First, two photons are distributed via dephasing channels to receiving parties, which converts the frequency-frequency entanglement to polarization-frequency correlations. This is followed by local up-conversion by each party removing the polarization-frequency correlations and which-path information in frequency space. Thereby the scheme ultimately allows for the remote creation of polarization entanglement.

We have identified two different mechanisms for this purpose depending on the initial form of the frequency distributions. When we use a two-peak structure for the local distributions, corresponding to non-Markovian dynamics, one can create polarization entanglement  both in short and long dephasing time regimes. If the frequency distribution has a single peak structure, corresponding to Markovian dephasing dynamics, one can create entanglement asymptotically. In the latter case, the stronger is the dephasing, the faster the asymptotic high-entanglement limit is reached. Moreover, the creation of polarization entanglement can be interpreted as destructive and constructive interference between the frequency mode paths in the up-conversion where only those polarization components that form the entangled state get up-converted. 

In addition of carrying both fundamental importance for quantum interferometry and practical importance for entanglement generation, our results also pave the way to the development of techniques to detect the  entanglement between the photon frequencies. When the initial amount of frequency entanglement is not known, our scheme allows us -- by measuring the amount of created polarization entanglement --  to infer whether there existed initial frequency entanglement or not. This is the key ingredient to implement a quantum probe for entanglement detection. We think that proof-of-principle experiment able to demonstrate remote polarization entanglement generation by local dephasing and frequency upconversion is currently within the grasp of the experimentalists.
 
\acknowledgements
This work has been supported by the EU through the Collaborative Projects QuProCS (Grant Agreement No. 641277)
and by the Academy of Finland (Project no. 287750). 
The USTC group acknowledges funding from the National Natural Science Foundation of China (Nos. 61327901,11325419), the Strategic Priority Research Program (B) of the Chinese Academy of Sciences (Grant No. XDB01030300), and Key Research Program of Frontier Sciences, CAS (No. QYZDY-SSW-SLH003).
G.K. is grateful to Sao Paulo Research Foundation (FAPESP)
for the support given under the grant number 2012/18558-5.
S.H.R. is grateful to CIMO for the support given under the grant number TM-15-9763.

%

\begin{thebibliography}{25}%
\makeatletter
\providecommand \@ifxundefined [1]{%
 \@ifx{#1\undefined}
}%
\providecommand \@ifnum [1]{%
 \ifnum #1\expandafter \@firstoftwo
 \else \expandafter \@secondoftwo
 \fi
}%
\providecommand \@ifx [1]{%
 \ifx #1\expandafter \@firstoftwo
 \else \expandafter \@secondoftwo
 \fi
}%
\providecommand \natexlab [1]{#1}%
\providecommand \enquote  [1]{``#1''}%
\providecommand \bibnamefont  [1]{#1}%
\providecommand \bibfnamefont [1]{#1}%
\providecommand \citenamefont [1]{#1}%
\providecommand \href@noop [0]{\@secondoftwo}%
\providecommand \href [0]{\begingroup \@sanitize@url \@href}%
\providecommand \@href[1]{\@@startlink{#1}\@@href}%
\providecommand \@@href[1]{\endgroup#1\@@endlink}%
\providecommand \@sanitize@url [0]{\catcode `\\12\catcode `\$12\catcode
  `\&12\catcode `\#12\catcode `\^12\catcode `\_12\catcode `\%12\relax}%
\providecommand \@@startlink[1]{}%
\providecommand \@@endlink[0]{}%
\providecommand \url  [0]{\begingroup\@sanitize@url \@url }%
\providecommand \@url [1]{\endgroup\@href {#1}{\urlprefix }}%
\providecommand \urlprefix  [0]{URL }%
\providecommand \Eprint [0]{\href }%
\providecommand \doibase [0]{http://dx.doi.org/}%
\providecommand \selectlanguage [0]{\@gobble}%
\providecommand \bibinfo  [0]{\@secondoftwo}%
\providecommand \bibfield  [0]{\@secondoftwo}%
\providecommand \translation [1]{[#1]}%
\providecommand \BibitemOpen [0]{}%
\providecommand \bibitemStop [0]{}%
\providecommand \bibitemNoStop [0]{.\EOS\space}%
\providecommand \EOS [0]{\spacefactor3000\relax}%
\providecommand \BibitemShut  [1]{\csname bibitem#1\endcsname}%
\let\auto@bib@innerbib\@empty
\bibitem [{\citenamefont {Ramelow}\ \emph {et~al.}(2009)\citenamefont
  {Ramelow}, \citenamefont {Ratschbacher}, \citenamefont {Fedrizzi},
  \citenamefont {Langford},\ and\ \citenamefont {Zeilinger}}]{zeilinger-2009}%
  \BibitemOpen
  \bibfield  {author} {\bibinfo {author} {\bibfnamefont {S.}~\bibnamefont
  {Ramelow}}, \bibinfo {author} {\bibfnamefont {L.}~\bibnamefont
  {Ratschbacher}}, \bibinfo {author} {\bibfnamefont {A.}~\bibnamefont
  {Fedrizzi}}, \bibinfo {author} {\bibfnamefont {N.~K.}\ \bibnamefont
  {Langford}}, \ and\ \bibinfo {author} {\bibfnamefont {A.}~\bibnamefont
  {Zeilinger}},\ }\href {\doibase 10.1103/PhysRevLett.103.253601} {\bibfield
  {journal} {\bibinfo  {journal} {Phys. Rev. Lett.}\ }\textbf {\bibinfo
  {volume} {103}},\ \bibinfo {pages} {253601} (\bibinfo {year}
  {2009})}\BibitemShut {NoStop}%
\bibitem [{\citenamefont {Zhao}\ \emph {et~al.}(2014)\citenamefont {Zhao},
  \citenamefont {Zhang}, \citenamefont {Yang}, \citenamefont {Sang},
  \citenamefont {Jiang}, \citenamefont {Bao},\ and\ \citenamefont
  {Pan}}]{zhao_entangling_2014}%
  \BibitemOpen
  \bibfield  {author} {\bibinfo {author} {\bibfnamefont {T.-M.}\ \bibnamefont
  {Zhao}}, \bibinfo {author} {\bibfnamefont {H.}~\bibnamefont {Zhang}},
  \bibinfo {author} {\bibfnamefont {J.}~\bibnamefont {Yang}}, \bibinfo {author}
  {\bibfnamefont {Z.-R.}\ \bibnamefont {Sang}}, \bibinfo {author}
  {\bibfnamefont {X.}~\bibnamefont {Jiang}}, \bibinfo {author} {\bibfnamefont
  {X.-H.}\ \bibnamefont {Bao}}, \ and\ \bibinfo {author} {\bibfnamefont
  {J.-W.}\ \bibnamefont {Pan}},\ }\href {\doibase
  10.1103/PhysRevLett.112.103602} {\bibfield  {journal} {\bibinfo  {journal}
  {Phys. Rev. Lett.}\ }\textbf {\bibinfo {volume} {112}},\ \bibinfo {pages}
  {103602} (\bibinfo {year} {2014})}\BibitemShut {NoStop}%
\bibitem [{\citenamefont {Cho}\ \emph {et~al.}(2014)\citenamefont {Cho},
  \citenamefont {Park}, \citenamefont {Lee},\ and\ \citenamefont
  {Kim}}]{cho_engineering_2014}%
  \BibitemOpen
  \bibfield  {author} {\bibinfo {author} {\bibfnamefont {Y.-W.}\ \bibnamefont
  {Cho}}, \bibinfo {author} {\bibfnamefont {K.-K.}\ \bibnamefont {Park}},
  \bibinfo {author} {\bibfnamefont {J.-C.}\ \bibnamefont {Lee}}, \ and\
  \bibinfo {author} {\bibfnamefont {Y.-H.}\ \bibnamefont {Kim}},\ }\href
  {\doibase 10.1103/PhysRevLett.113.063602} {\bibfield  {journal} {\bibinfo
  {journal} {Phys. Rev. Lett.}\ }\textbf {\bibinfo {volume} {113}},\ \bibinfo
  {pages} {063602} (\bibinfo {year} {2014})}\BibitemShut {NoStop}%
\bibitem [{\citenamefont {Roslund}\ \emph {et~al.}(2014)\citenamefont
  {Roslund}, \citenamefont {de~Ara{\'u}jo}, \citenamefont {Jiang},
  \citenamefont {Fabre},\ and\ \citenamefont
  {Treps}}]{roslund_wavelength-multiplexed_2014-1}%
  \BibitemOpen
  \bibfield  {author} {\bibinfo {author} {\bibfnamefont {J.}~\bibnamefont
  {Roslund}}, \bibinfo {author} {\bibfnamefont {R.~M.}\ \bibnamefont
  {de~Ara{\'u}jo}}, \bibinfo {author} {\bibfnamefont {S.}~\bibnamefont
  {Jiang}}, \bibinfo {author} {\bibfnamefont {C.}~\bibnamefont {Fabre}}, \ and\
  \bibinfo {author} {\bibfnamefont {N.}~\bibnamefont {Treps}},\ }\href
  {\doibase 10.1038/nphoton.2013.340} {\bibfield  {journal} {\bibinfo
  {journal} {Nature Photon.}\ }\textbf {\bibinfo {volume} {8}},\ \bibinfo
  {pages} {109} (\bibinfo {year} {2014})}\BibitemShut {NoStop}%
\bibitem [{\citenamefont {Reimer}\ \emph {et~al.}(2016)\citenamefont {Reimer},
  \citenamefont {Kues}, \citenamefont {Roztocki}, \citenamefont {Wetzel},
  \citenamefont {Grazioso}, \citenamefont {Little}, \citenamefont {Chu},
  \citenamefont {Johnston}, \citenamefont {Bromberg}, \citenamefont {Caspani},
  \citenamefont {Moss},\ and\ \citenamefont {Morandotti}}]{Reimer1176}%
  \BibitemOpen
  \bibfield  {author} {\bibinfo {author} {\bibfnamefont {C.}~\bibnamefont
  {Reimer}}, \bibinfo {author} {\bibfnamefont {M.}~\bibnamefont {Kues}},
  \bibinfo {author} {\bibfnamefont {P.}~\bibnamefont {Roztocki}}, \bibinfo
  {author} {\bibfnamefont {B.}~\bibnamefont {Wetzel}}, \bibinfo {author}
  {\bibfnamefont {F.}~\bibnamefont {Grazioso}}, \bibinfo {author}
  {\bibfnamefont {B.~E.}\ \bibnamefont {Little}}, \bibinfo {author}
  {\bibfnamefont {S.~T.}\ \bibnamefont {Chu}}, \bibinfo {author} {\bibfnamefont
  {T.}~\bibnamefont {Johnston}}, \bibinfo {author} {\bibfnamefont
  {Y.}~\bibnamefont {Bromberg}}, \bibinfo {author} {\bibfnamefont
  {L.}~\bibnamefont {Caspani}}, \bibinfo {author} {\bibfnamefont {D.~J.}\
  \bibnamefont {Moss}}, \ and\ \bibinfo {author} {\bibfnamefont
  {R.}~\bibnamefont {Morandotti}},\ }\href {\doibase 10.1126/science.aad8532}
  {\bibfield  {journal} {\bibinfo  {journal} {Science}\ }\textbf {\bibinfo
  {volume} {351}},\ \bibinfo {pages} {1176} (\bibinfo {year}
  {2016})}\BibitemShut {NoStop}%
\bibitem [{\citenamefont {Kobayashi}\ \emph {et~al.}(2016)\citenamefont
  {Kobayashi}, \citenamefont {Ikuta}, \citenamefont {Yasui}, \citenamefont
  {Miki}, \citenamefont {Yamashita}, \citenamefont {Terai}, \citenamefont
  {Yamamoto}, \citenamefont {Koashi},\ and\ \citenamefont
  {Imoto}}]{kobayashi_frequency-domain_2016}%
  \BibitemOpen
  \bibfield  {author} {\bibinfo {author} {\bibfnamefont {T.}~\bibnamefont
  {Kobayashi}}, \bibinfo {author} {\bibfnamefont {R.}~\bibnamefont {Ikuta}},
  \bibinfo {author} {\bibfnamefont {S.}~\bibnamefont {Yasui}}, \bibinfo
  {author} {\bibfnamefont {S.}~\bibnamefont {Miki}}, \bibinfo {author}
  {\bibfnamefont {T.}~\bibnamefont {Yamashita}}, \bibinfo {author}
  {\bibfnamefont {H.}~\bibnamefont {Terai}}, \bibinfo {author} {\bibfnamefont
  {T.}~\bibnamefont {Yamamoto}}, \bibinfo {author} {\bibfnamefont
  {M.}~\bibnamefont {Koashi}}, \ and\ \bibinfo {author} {\bibfnamefont
  {N.}~\bibnamefont {Imoto}},\ }\href {\doibase 10.1038/nphoton.2016.74}
  {\bibfield  {journal} {\bibinfo  {journal} {Nature Photon.}\ }\textbf
  {\bibinfo {volume} {10}},\ \bibinfo {pages} {441} (\bibinfo {year}
  {2016})}\BibitemShut {NoStop}%
\bibitem [{\citenamefont {Aolita}\ and\ \citenamefont
  {Walborn}(2007)}]{aolita_quantum_2007}%
  \BibitemOpen
  \bibfield  {author} {\bibinfo {author} {\bibfnamefont {L.}~\bibnamefont
  {Aolita}}\ and\ \bibinfo {author} {\bibfnamefont {S.~P.}\ \bibnamefont
  {Walborn}},\ }\href {\doibase 10.1103/PhysRevLett.98.100501} {\bibfield
  {journal} {\bibinfo  {journal} {Phys. Rev. Lett.}\ }\textbf {\bibinfo
  {volume} {98}},\ \bibinfo {pages} {100501} (\bibinfo {year}
  {2007})}\BibitemShut {NoStop}%
\bibitem [{\citenamefont {Xiao}\ \emph {et~al.}(2008)\citenamefont {Xiao},
  \citenamefont {Wang}, \citenamefont {Zhang}, \citenamefont {Huang},
  \citenamefont {Peng},\ and\ \citenamefont {Long}}]{xiao_efficient_2008}%
  \BibitemOpen
  \bibfield  {author} {\bibinfo {author} {\bibfnamefont {L.}~\bibnamefont
  {Xiao}}, \bibinfo {author} {\bibfnamefont {C.}~\bibnamefont {Wang}}, \bibinfo
  {author} {\bibfnamefont {W.}~\bibnamefont {Zhang}}, \bibinfo {author}
  {\bibfnamefont {Y.}~\bibnamefont {Huang}}, \bibinfo {author} {\bibfnamefont
  {J.}~\bibnamefont {Peng}}, \ and\ \bibinfo {author} {\bibfnamefont
  {G.}~\bibnamefont {Long}},\ }\href {\doibase 10.1103/PhysRevA.77.042315}
  {\bibfield  {journal} {\bibinfo  {journal} {Phys. Rev. A.}\ }\textbf
  {\bibinfo {volume} {77}},\ \bibinfo {pages} {042315} (\bibinfo {year}
  {2008})}\BibitemShut {NoStop}%
\bibitem [{\citenamefont {Wang}\ \emph {et~al.}(2009)\citenamefont {Wang},
  \citenamefont {Xiao}, \citenamefont {Wang}, \citenamefont {Zhang},\ and\
  \citenamefont {Long}}]{wang_quantum_2009}%
  \BibitemOpen
  \bibfield  {author} {\bibinfo {author} {\bibfnamefont {C.}~\bibnamefont
  {Wang}}, \bibinfo {author} {\bibfnamefont {L.}~\bibnamefont {Xiao}}, \bibinfo
  {author} {\bibfnamefont {W.-y.}\ \bibnamefont {Wang}}, \bibinfo {author}
  {\bibfnamefont {G.-y.}\ \bibnamefont {Zhang}}, \ and\ \bibinfo {author}
  {\bibfnamefont {G.~L.}\ \bibnamefont {Long}},\ }\href {\doibase
  10.1364/JOSAB.26.002072} {\bibfield  {journal} {\bibinfo  {journal} {J. Opt.
  Soc. Am. B}\ }\textbf {\bibinfo {volume} {26}},\ \bibinfo {pages} {2072}
  (\bibinfo {year} {2009})}\BibitemShut {NoStop}%
\bibitem [{\citenamefont {De~Greve}\ \emph {et~al.}(2012)\citenamefont
  {De~Greve}, \citenamefont {Yu}, \citenamefont {McMahon}, \citenamefont
  {Pelc}, \citenamefont {Natarajan}, \citenamefont {Kim}, \citenamefont {Abe},
  \citenamefont {Maier}, \citenamefont {Schneider}, \citenamefont {Kamp},
  \citenamefont {Hofling}, \citenamefont {Hadfield}, \citenamefont {Forchel},
  \citenamefont {Fejer},\ and\ \citenamefont {Yamamoto}}]{degreve-2012}%
  \BibitemOpen
  \bibfield  {author} {\bibinfo {author} {\bibfnamefont {K.}~\bibnamefont
  {De~Greve}}, \bibinfo {author} {\bibfnamefont {L.}~\bibnamefont {Yu}},
  \bibinfo {author} {\bibfnamefont {P.~L.}\ \bibnamefont {McMahon}}, \bibinfo
  {author} {\bibfnamefont {J.~S.}\ \bibnamefont {Pelc}}, \bibinfo {author}
  {\bibfnamefont {C.~M.}\ \bibnamefont {Natarajan}}, \bibinfo {author}
  {\bibfnamefont {N.~Y.}\ \bibnamefont {Kim}}, \bibinfo {author} {\bibfnamefont
  {E.}~\bibnamefont {Abe}}, \bibinfo {author} {\bibfnamefont {S.}~\bibnamefont
  {Maier}}, \bibinfo {author} {\bibfnamefont {C.}~\bibnamefont {Schneider}},
  \bibinfo {author} {\bibfnamefont {M.}~\bibnamefont {Kamp}}, \bibinfo {author}
  {\bibfnamefont {S.}~\bibnamefont {Hofling}}, \bibinfo {author} {\bibfnamefont
  {R.~H.}\ \bibnamefont {Hadfield}}, \bibinfo {author} {\bibfnamefont
  {A.}~\bibnamefont {Forchel}}, \bibinfo {author} {\bibfnamefont {M.~M.}\
  \bibnamefont {Fejer}}, \ and\ \bibinfo {author} {\bibfnamefont
  {Y.}~\bibnamefont {Yamamoto}},\ }\href {\doibase 10.1038/nature11577}
  {\bibfield  {journal} {\bibinfo  {journal} {Nature}\ }\textbf {\bibinfo
  {volume} {491}},\ \bibinfo {pages} {421} (\bibinfo {year}
  {2012})}\BibitemShut {NoStop}%
\bibitem [{\citenamefont {Nagali}\ \emph {et~al.}(2009)\citenamefont {Nagali},
  \citenamefont {Sciarrino}, \citenamefont {De~Martini}, \citenamefont
  {Marrucci}, \citenamefont {Piccirillo}, \citenamefont {Karimi},\ and\
  \citenamefont {Santamato}}]{nagali_quantum_2009}%
  \BibitemOpen
  \bibfield  {author} {\bibinfo {author} {\bibfnamefont {E.}~\bibnamefont
  {Nagali}}, \bibinfo {author} {\bibfnamefont {F.}~\bibnamefont {Sciarrino}},
  \bibinfo {author} {\bibfnamefont {F.}~\bibnamefont {De~Martini}}, \bibinfo
  {author} {\bibfnamefont {L.}~\bibnamefont {Marrucci}}, \bibinfo {author}
  {\bibfnamefont {B.}~\bibnamefont {Piccirillo}}, \bibinfo {author}
  {\bibfnamefont {E.}~\bibnamefont {Karimi}}, \ and\ \bibinfo {author}
  {\bibfnamefont {E.}~\bibnamefont {Santamato}},\ }\href {\doibase
  10.1103/PhysRevLett.103.013601} {\bibfield  {journal} {\bibinfo  {journal}
  {Phys. Rev. Lett.}\ }\textbf {\bibinfo {volume} {103}},\ \bibinfo {pages}
  {013601} (\bibinfo {year} {2009})}\BibitemShut {NoStop}%
\bibitem [{\citenamefont {Takesue}(2008)}]{takesue-2008}%
  \BibitemOpen
  \bibfield  {author} {\bibinfo {author} {\bibfnamefont {H.}~\bibnamefont
  {Takesue}},\ }\href {\doibase 10.1103/PhysRevLett.101.173901} {\bibfield
  {journal} {\bibinfo  {journal} {Phys. Rev. Lett.}\ }\textbf {\bibinfo
  {volume} {101}},\ \bibinfo {pages} {173901} (\bibinfo {year}
  {2008})}\BibitemShut {NoStop}%
\bibitem [{\citenamefont {Tanzilli}\ \emph {et~al.}(2005)\citenamefont
  {Tanzilli}, \citenamefont {Tittel}, \citenamefont {Halder}, \citenamefont
  {Alibart}, \citenamefont {Baldi}, \citenamefont {Gisin},\ and\ \citenamefont
  {Zbinden}}]{Tanzilli:2005aa}%
  \BibitemOpen
  \bibfield  {author} {\bibinfo {author} {\bibfnamefont {S.}~\bibnamefont
  {Tanzilli}}, \bibinfo {author} {\bibfnamefont {W.}~\bibnamefont {Tittel}},
  \bibinfo {author} {\bibfnamefont {M.}~\bibnamefont {Halder}}, \bibinfo
  {author} {\bibfnamefont {O.}~\bibnamefont {Alibart}}, \bibinfo {author}
  {\bibfnamefont {P.}~\bibnamefont {Baldi}}, \bibinfo {author} {\bibfnamefont
  {N.}~\bibnamefont {Gisin}}, \ and\ \bibinfo {author} {\bibfnamefont
  {H.}~\bibnamefont {Zbinden}},\ }\href {http://dx.doi.org/10.1038/nature04009}
  {\bibfield  {journal} {\bibinfo  {journal} {Nature}\ }\textbf {\bibinfo
  {volume} {437}},\ \bibinfo {pages} {116} (\bibinfo {year}
  {2005})}\BibitemShut {NoStop}%
\bibitem [{\citenamefont {Ikuta}\ \emph {et~al.}(2011)\citenamefont {Ikuta},
  \citenamefont {Kusaka}, \citenamefont {Kitano}, \citenamefont {Kato},
  \citenamefont {Yamamoto}, \citenamefont {Koashi},\ and\ \citenamefont
  {Imoto}}]{Ikuta:2011aa}%
  \BibitemOpen
  \bibfield  {author} {\bibinfo {author} {\bibfnamefont {R.}~\bibnamefont
  {Ikuta}}, \bibinfo {author} {\bibfnamefont {Y.}~\bibnamefont {Kusaka}},
  \bibinfo {author} {\bibfnamefont {T.}~\bibnamefont {Kitano}}, \bibinfo
  {author} {\bibfnamefont {H.}~\bibnamefont {Kato}}, \bibinfo {author}
  {\bibfnamefont {T.}~\bibnamefont {Yamamoto}}, \bibinfo {author}
  {\bibfnamefont {M.}~\bibnamefont {Koashi}}, \ and\ \bibinfo {author}
  {\bibfnamefont {N.}~\bibnamefont {Imoto}},\ }\href
  {http://dx.doi.org/10.1038/ncomms1544} {\bibfield  {journal} {\bibinfo
  {journal} {Nature Comm.}\ }\textbf {\bibinfo {volume} {2}},\ \bibinfo {pages}
  {1544} (\bibinfo {year} {2011})}\BibitemShut {NoStop}%
\bibitem [{\citenamefont {Xu}\ \emph {et~al.}(2010)\citenamefont {Xu},
  \citenamefont {Xu}, \citenamefont {Li}, \citenamefont {Zhang}, \citenamefont
  {Zou},\ and\ \citenamefont {Guo}}]{feng-2010}%
  \BibitemOpen
  \bibfield  {author} {\bibinfo {author} {\bibfnamefont {J.-S.}\ \bibnamefont
  {Xu}}, \bibinfo {author} {\bibfnamefont {X.-Y.}\ \bibnamefont {Xu}}, \bibinfo
  {author} {\bibfnamefont {C.-F.}\ \bibnamefont {Li}}, \bibinfo {author}
  {\bibfnamefont {C.-J.}\ \bibnamefont {Zhang}}, \bibinfo {author}
  {\bibfnamefont {X.-B.}\ \bibnamefont {Zou}}, \ and\ \bibinfo {author}
  {\bibfnamefont {G.-C.}\ \bibnamefont {Guo}},\ }\href
  {http://dx.doi.org/10.1038/ncomms1005} {\bibfield  {journal} {\bibinfo
  {journal} {Nature Comm.}\ }\textbf {\bibinfo {volume} {1}},\ \bibinfo {pages}
  {7} (\bibinfo {year} {2010})}\BibitemShut {NoStop}%
\bibitem [{\citenamefont {Kwiat}\ \emph {et~al.}(2000)\citenamefont {Kwiat},
  \citenamefont {Berglund}, \citenamefont {Altepeter},\ and\ \citenamefont
  {White}}]{kwiat-2000}%
  \BibitemOpen
  \bibfield  {author} {\bibinfo {author} {\bibfnamefont {P.~G.}\ \bibnamefont
  {Kwiat}}, \bibinfo {author} {\bibfnamefont {A.~J.}\ \bibnamefont {Berglund}},
  \bibinfo {author} {\bibfnamefont {J.~B.}\ \bibnamefont {Altepeter}}, \ and\
  \bibinfo {author} {\bibfnamefont {A.~G.}\ \bibnamefont {White}},\ }\href
  {\doibase 10.1126/science.290.5491.498} {\bibfield  {journal} {\bibinfo
  {journal} {Science}\ }\textbf {\bibinfo {volume} {290}},\ \bibinfo {pages}
  {498} (\bibinfo {year} {2000})}\BibitemShut {NoStop}%
\bibitem [{\citenamefont {Laine}\ \emph {et~al.}(2012)\citenamefont {Laine},
  \citenamefont {Breuer}, \citenamefont {Piilo}, \citenamefont {Li},\ and\
  \citenamefont {Guo}}]{Laine2012a}%
  \BibitemOpen
  \bibfield  {author} {\bibinfo {author} {\bibfnamefont {E.-M.}\ \bibnamefont
  {Laine}}, \bibinfo {author} {\bibfnamefont {H.-P.}\ \bibnamefont {Breuer}},
  \bibinfo {author} {\bibfnamefont {J.}~\bibnamefont {Piilo}}, \bibinfo
  {author} {\bibfnamefont {C.-F.}\ \bibnamefont {Li}}, \ and\ \bibinfo {author}
  {\bibfnamefont {G.-C.}\ \bibnamefont {Guo}},\ }\href {\doibase
  10.1103/PhysRevLett.108.210402} {\bibfield  {journal} {\bibinfo  {journal}
  {Phys. Rev. Lett.}\ }\textbf {\bibinfo {volume} {108}},\ \bibinfo {pages}
  {210402} (\bibinfo {year} {2012})}\BibitemShut {NoStop}%
\bibitem [{\citenamefont {Laine}\ \emph {et~al.}(2013)\citenamefont {Laine},
  \citenamefont {Breuer}, \citenamefont {Piilo}, \citenamefont {Li},\ and\
  \citenamefont {Guo}}]{Laine2013b}%
  \BibitemOpen
  \bibfield  {author} {\bibinfo {author} {\bibfnamefont {E.-M.}\ \bibnamefont
  {Laine}}, \bibinfo {author} {\bibfnamefont {H.-P.}\ \bibnamefont {Breuer}},
  \bibinfo {author} {\bibfnamefont {J.}~\bibnamefont {Piilo}}, \bibinfo
  {author} {\bibfnamefont {C.-F.}\ \bibnamefont {Li}}, \ and\ \bibinfo {author}
  {\bibfnamefont {G.-C.}\ \bibnamefont {Guo}},\ }\href {\doibase
  10.1103/PhysRevLett.111.229901} {\bibfield  {journal} {\bibinfo  {journal}
  {Phys. Rev. Lett.}\ }\textbf {\bibinfo {volume} {111}},\ \bibinfo {pages}
  {229901} (\bibinfo {year} {2013})}\BibitemShut {NoStop}%
\bibitem [{\citenamefont {Liu}\ \emph {et~al.}(2013)\citenamefont {Liu},
  \citenamefont {Cao}, \citenamefont {Huang}, \citenamefont {Li}, \citenamefont
  {Guo}, \citenamefont {Laine}, \citenamefont {Breuer},\ and\ \citenamefont
  {Piilo}}]{Liu2013a}%
  \BibitemOpen
  \bibfield  {author} {\bibinfo {author} {\bibfnamefont {B.-H.}\ \bibnamefont
  {Liu}}, \bibinfo {author} {\bibfnamefont {D.-Y.}\ \bibnamefont {Cao}},
  \bibinfo {author} {\bibfnamefont {Y.-F.}\ \bibnamefont {Huang}}, \bibinfo
  {author} {\bibfnamefont {C.-F.}\ \bibnamefont {Li}}, \bibinfo {author}
  {\bibfnamefont {G.-C.}\ \bibnamefont {Guo}}, \bibinfo {author} {\bibfnamefont
  {E.-M.}\ \bibnamefont {Laine}}, \bibinfo {author} {\bibfnamefont {H.-P.}\
  \bibnamefont {Breuer}}, \ and\ \bibinfo {author} {\bibfnamefont
  {J.}~\bibnamefont {Piilo}},\ }\href {http://dx.doi.org/10.1038/srep01781}
  {\bibfield  {journal} {\bibinfo  {journal} {Sci. Rep.}\ }\textbf {\bibinfo
  {volume} {3}} (\bibinfo {year} {2013})}\BibitemShut {NoStop}%
\bibitem [{\citenamefont {Scully}\ and\ \citenamefont
  {Dr\"uhl}(1982)}]{scully-1982}%
  \BibitemOpen
  \bibfield  {author} {\bibinfo {author} {\bibfnamefont {M.~O.}\ \bibnamefont
  {Scully}}\ and\ \bibinfo {author} {\bibfnamefont {K.}~\bibnamefont
  {Dr\"uhl}},\ }\href {\doibase 10.1103/PhysRevA.25.2208} {\bibfield  {journal}
  {\bibinfo  {journal} {Phys. Rev. A}\ }\textbf {\bibinfo {volume} {25}},\
  \bibinfo {pages} {2208} (\bibinfo {year} {1982})}\BibitemShut {NoStop}%
\bibitem [{\citenamefont {Garisto}\ and\ \citenamefont
  {Hardy}(1999)}]{PhysRevA.60.827}%
  \BibitemOpen
  \bibfield  {author} {\bibinfo {author} {\bibfnamefont {R.}~\bibnamefont
  {Garisto}}\ and\ \bibinfo {author} {\bibfnamefont {L.}~\bibnamefont
  {Hardy}},\ }\href {\doibase 10.1103/PhysRevA.60.827} {\bibfield  {journal}
  {\bibinfo  {journal} {Phys. Rev. A}\ }\textbf {\bibinfo {volume} {60}},\
  \bibinfo {pages} {827} (\bibinfo {year} {1999})}\BibitemShut {NoStop}%
\bibitem [{\citenamefont {Liu}\ \emph {et~al.}(2011)\citenamefont {Liu},
  \citenamefont {Li}, \citenamefont {Huang}, \citenamefont {Li}, \citenamefont
  {Guo}, \citenamefont {Laine}, \citenamefont {Breuer},\ and\ \citenamefont
  {Piilo}}]{Liu:2011aa}%
  \BibitemOpen
  \bibfield  {author} {\bibinfo {author} {\bibfnamefont {B.-H.}\ \bibnamefont
  {Liu}}, \bibinfo {author} {\bibfnamefont {L.}~\bibnamefont {Li}}, \bibinfo
  {author} {\bibfnamefont {Y.-F.}\ \bibnamefont {Huang}}, \bibinfo {author}
  {\bibfnamefont {C.-F.}\ \bibnamefont {Li}}, \bibinfo {author} {\bibfnamefont
  {G.-C.}\ \bibnamefont {Guo}}, \bibinfo {author} {\bibfnamefont {E.-M.}\
  \bibnamefont {Laine}}, \bibinfo {author} {\bibfnamefont {H.-P.}\ \bibnamefont
  {Breuer}}, \ and\ \bibinfo {author} {\bibfnamefont {J.}~\bibnamefont
  {Piilo}},\ }\href {http://dx.doi.org/10.1038/nphys2085} {\bibfield  {journal}
  {\bibinfo  {journal} {Nature Phys.}\ }\textbf {\bibinfo {volume} {7}},\
  \bibinfo {pages} {931} (\bibinfo {year} {2011})}\BibitemShut {NoStop}%
\bibitem [{\citenamefont {Ou}\ \emph {et~al.}(1999)\citenamefont {Ou},
  \citenamefont {Rhee},\ and\ \citenamefont {Wang}}]{Ou99}%
  \BibitemOpen
  \bibfield  {author} {\bibinfo {author} {\bibfnamefont {Z.~Y.}\ \bibnamefont
  {Ou}}, \bibinfo {author} {\bibfnamefont {J.-K.}\ \bibnamefont {Rhee}}, \ and\
  \bibinfo {author} {\bibfnamefont {L.~J.}\ \bibnamefont {Wang}},\ }\href@noop
  {} {\bibfield  {journal} {\bibinfo  {journal} {Phys. Rev A}\ }\textbf
  {\bibinfo {volume} {60}},\ \bibinfo {pages} {593} (\bibinfo {year}
  {1999})}\BibitemShut {NoStop}%
\bibitem [{\citenamefont {Wootters}(1998)}]{concurrence}%
  \BibitemOpen
  \bibfield  {author} {\bibinfo {author} {\bibfnamefont {W.~K.}\ \bibnamefont
  {Wootters}},\ }\href {\doibase 10.1103/PhysRevLett.80.2245} {\bibfield
  {journal} {\bibinfo  {journal} {Phys. Rev. Lett.}\ }\textbf {\bibinfo
  {volume} {80}},\ \bibinfo {pages} {2245} (\bibinfo {year}
  {1998})}\BibitemShut {NoStop}%
\bibitem [{\citenamefont {Liu}\ \emph {et~al.}(2016)\citenamefont {Liu},
  \citenamefont {Hu}, \citenamefont {Huang}, \citenamefont {Li}, \citenamefont
  {Guo}, \citenamefont {Karlsson}, \citenamefont {Laine}, \citenamefont
  {Maniscalco}, \citenamefont {Macchiavello},\ and\ \citenamefont
  {Piilo}}]{Liu2016a}%
  \BibitemOpen
  \bibfield  {author} {\bibinfo {author} {\bibfnamefont {B.-H.}\ \bibnamefont
  {Liu}}, \bibinfo {author} {\bibfnamefont {Z.-M.}\ \bibnamefont {Hu}},
  \bibinfo {author} {\bibfnamefont {Y.-F.}\ \bibnamefont {Huang}}, \bibinfo
  {author} {\bibfnamefont {C.-F.}\ \bibnamefont {Li}}, \bibinfo {author}
  {\bibfnamefont {G.-C.}\ \bibnamefont {Guo}}, \bibinfo {author} {\bibfnamefont
  {A.}~\bibnamefont {Karlsson}}, \bibinfo {author} {\bibfnamefont {E.-M.}\
  \bibnamefont {Laine}}, \bibinfo {author} {\bibfnamefont {S.}~\bibnamefont
  {Maniscalco}}, \bibinfo {author} {\bibfnamefont {C.}~\bibnamefont
  {Macchiavello}}, \ and\ \bibinfo {author} {\bibfnamefont {J.}~\bibnamefont
  {Piilo}},\ }\href@noop {} {\bibfield  {journal} {\bibinfo  {journal} {EPL
  (Europhysics Letters)}\ }\textbf {\bibinfo {volume} {114}},\ \bibinfo {pages}
  {10005} (\bibinfo {year} {2016})}\BibitemShut {NoStop}%
\end{thebibliography}

\end{document}